\documentclass[aps,prb,twocolumn,superscriptaddress,showpacs,floatfix]{revtex4}
\usepackage{graphicx}
\usepackage{amsmath} 

\begin{document}

\title{Stand form of the scattering matrix for time reversal symmetric system}

\author{Yongjin Jiang}\email{jyj@zjnu.cn}
\author{Xiaoli Lu}
\author{Feng Zhai}
\affiliation{Center for Statistical and Theoretical Condensed
Matter Physics, and Department of Physics, Zhejiang Normal
University, Jinhua 321004, People's Republic of China}
\begin{abstract}
  In this paper, we present the standard form of the scattering
matrix of mesocopic system with spin-orbital coupling
which preserves time reversal symmetry. We found some analytical structure  of the scattering matrix related to the sub-matrices between arbitrary two channels. In particular, we proved
that in the two-terminal mono-channel scattering problem, the transmission matrix is proportional to a $SU(2)$
matrix. We  obtained these properties  through direct and elementary way and found it in agreement with polar decomposition  known before.
\end{abstract}

 \maketitle


Time reversal symmetry(TRS) is a general symmetry in various
physical systems. In quantum mechanics, it has exceptionally deep
consequences. In particular, the  Kramer's degeneracy, i.e.,
double degeneracy of energy eigenvalues, for spin 1/2 particle
system with TRS is well-known since the early days of quantum
mechanics\cite{Kramers}. The transmission eigenvalues of a two-terminal TRS system  has also
similar double-degeneracy(for nonzero transmission channels)\cite{Bardarson}. Such
degeneracy results from antisymmetric property of the scattering
matrix when the incoming states and the out-going states are
properly ordered. In a recent
paper\cite{Jiang1}, we obtained several theorems on the
 connectivity property of edge states in topological
insulators\cite{Qizhang,HasanKane} in which the antisymmetry
property of the scattering matrix is further explored.

 On the
other hand, the scattering matrix in the usually ordered basis has
another interesting property\cite{Bardarson,Beenakker},

\begin{eqnarray}
\sigma_yS\sigma_y=S^{T}   \label{eq:8}
\end{eqnarray}
for which we will provide an explicit derivation in the following.
 In this paper, we will derive the standard form of $S$ from the restriction Eq.(\ref{eq:8}). In particular, we found that in a
 two-terminal scattering problem with single channel leads,  the transmission
 matrix is proportional to a $SU(2)$ matrix. These results will be
 useful in theoretical analysis of spin-related effects such as
 spin pumping effect\cite{Nagaosa}, topological
 classification\cite{MeidanBrower,Akhmerow}, etc.

Let's consider an one-dimensional system, which is composed of a
central region with spin-orbital coupling and two ideal metallic
leads(which are called left and right lead hereafter). For certain
energy $E$, the electronic wave function in the two leads is some
linear combinations of the eigenstates of the form
$|d\sigma\rangle$, where $d$ denotes the velocity direction, i.e.,
$d=1$ represents right-moving and $d=-1$ left moving, and $\sigma$
means the spin quantum number, i.e.,  $\sigma$ can be 1($\uparrow$) or -1($\downarrow$) in units
of $\frac{\hbar}{2}$ relative to the spin z-axis 
fixed before. Generally, the scattering wave functions $|\psi\rangle$ on the left
and right lead take the following form:
\begin{eqnarray}
|\psi\rangle_{L\sigma}=\phi^{in}_{L\sigma}|1\sigma\rangle+\phi^{out}_{L\sigma}|-1\sigma\rangle
\\
|\psi\rangle_{R\sigma}=\phi^{out}_{R\sigma}|1\sigma\rangle+\phi^{in}_{R\sigma}|-1\sigma\rangle
\end{eqnarray}
where the subscript $L(R)$ denotes left(right) lead, and
superscript $in(out)$ denotes incident(out-going)waves,
respectively. Let's define
$a=(\phi^{in}_{L\uparrow},\phi^{in}_{L\downarrow},\phi^{in}_{R\uparrow},\phi^{in}_{R\downarrow})^T$
as the incident wave amplitude and
$b=(\phi^{out}_{L\uparrow},\phi^{out}_{L\downarrow},\phi^{out}_{R\uparrow},\phi^{out}_{R\downarrow})^T$
as the out-scattering wave amplitude vector. By definition, the scattering
matrix $S$ satisfy:
\begin{eqnarray}
b=Sa \label{eq:3}
\end{eqnarray}
Taking complex conjugate, we have,
\begin{eqnarray}
a^*=(S)^{-1*}b^*=S^{T}b^*  \label{eq:4}
\end{eqnarray}
where we have used the unitarity property of S. Generally, we can
write the scattering matrix in the following form:
\begin{eqnarray}
S=\left(\begin{array}{cc}
  R & T' \\
  T & R' \\
\end{array}\right)\label{general0}
\end{eqnarray}

in which each entry $R$,$T$, etc is a $2\times2$ matrix. In the
usual terminology of scattering problem, $R$ and $R'$ are called
reflection matrix, while $T$ and $T'$ are transmission matrix.
Now, let's recall that for spin 1/2 particles, the time reversal
operator can be written as $\Theta=-i\sigma_yK$, where
$\sigma_y=\left(\begin{array}{cc}
  0 & -i \\
  i& 0 \\
\end{array}\right)$ is the standard Pauli matrix and $K$ is the
complex conjugate, which changes the direction of velocity, i.e.,
$K|\pm1\sigma\rangle=|\mp1\sigma\rangle$. Thus, $\Theta$ transforms the basis as follows, 
\begin{eqnarray}
\Theta=|\pm 1\sigma\rangle=\sigma|\mp 1\bar{\sigma}\rangle \label{eq7}
\end{eqnarray}
where $\bar{\sigma}=-\sigma$. In the following we call the basis satisfying Eq.(\ref{eq7}) as  normal basis.
From Eq.(\ref{eq:3}), we
have,
\begin{eqnarray}
\Theta b=\Theta S\Theta^{-1}\Theta a
\end{eqnarray}
where $\Theta b=-i\sigma_ya^*$ and $\Theta a=-i\sigma_yb^*$(note $\Theta a$($\Theta b$) is
the incident(out-going) wave amplitude vector in the time reversed frame).
Due to time reversal symmetry, we have $\Theta S\Theta ^{-1}=S$, so we have:
\begin{eqnarray}
\sigma_ya^*=S\sigma_yb^*
\end{eqnarray}
which, in combination with Eq.(\ref{eq:4}), results in
Eq.(\ref{eq:8}).
By changing to different basis, the scattering matrix can be
manifestly antisymmetric\cite{Bardarson,Jiang1}. In the following,
we will further discuss the constraint on transmission matrix due
to time reversal symmetry. Firstly, we can expand Eq.(\ref{eq:8})
in the following left/right block form:
\begin{eqnarray}
(\sigma_y)_{\alpha,\alpha'}(S)_{L\alpha',L\beta'}(\sigma_y)_{\beta',\beta}=(S)_{L\beta,L\alpha} \nonumber\\
(\sigma_y)_{\alpha,\alpha'}(S)_{L\alpha',R\beta'}(\sigma_y)_{\beta',\beta}=(S)_{R\beta,L\alpha} \label{eq:9}\\
(\sigma_y)_{\alpha,\alpha'}(S)_{R\alpha',R\beta'}(\sigma_y)_{\beta',\beta}=(S)_{R\beta,R\alpha}
\nonumber
\end{eqnarray}
where the matrix indices $\alpha,\beta$ take integer value 1(2)
for $\uparrow(\downarrow)$ spin states. By expressing the above
relations by reflection and transmission matrices, we obtain the
following form:
\begin{eqnarray}
&R&=rI,R'=r'I  \label{eq:10} \\
&T&=\left(\begin{array}{cc}
  t_1 & t_3 \\
  t_4& t_2 \\
\end{array}\right),T'=\left(\begin{array}{cc}
  t_2 & -t_3 \\
  -t_4& t_1 \\
\end{array}\right)=adj(T)\label{eq:11}
\end{eqnarray}
where $r,r',t_{i},i=1,...,4$ are unknown complex numbers and $I$
is the $2\times2$ unit matrix. $adj(T)$ is the adjugate matrix of $T$ so that $TT'=\det(T)I$.   From Eq.(\ref{eq:10}), it's evident
that an incident particle in the up spin state can't be reflected
to a down spin state(the incident and final modes being time
reversal pair). This property is discussed within the
circumstances of helical edge states of topological
insulator\cite{Qizhang,Jiang1} and  has profound physical
consequences. In what follows we will focus on the analytic property of the transmission matrix $T$. Obviously, both the determinant and trace of the
transmission matrices $T$ and $T'$ equal to each other:
\begin{eqnarray}
\det(T)=\det(T'), Tr(T)=Tr(T')
\end{eqnarray}
while the product of $T$ and $T'$ is,
\begin{eqnarray}
TT'=\left(\begin{array}{cc}
  t_1t_2-t_3t_4 & 0 \\
  0& t_1t_2-t_3t_4 \\
\end{array}\right)=\det(T)I
\end{eqnarray}
Once $\det(T)\not=0$, we have the general
form,
\begin{eqnarray}
T=tU, T'=tU^{-1}
\end{eqnarray}
where $t$ is determined by $t=\pm\sqrt{ \det(T)}$ up to a minus sign. Since
$\det(T)=\det(T')$, we have $\det(U)=\det(U^{-1})$, so that
$\det(U)=\pm 1$. Furthermore, from $Tr(T)=Tr(T')$, we have
$Tr(U)=Tr(U^{-1})$. With the same determinant and trace, it's
clear that the two-dimensional matrices $U$ and $U^{-1}$ has the
same eigenvalues.

Inputting the form Eq.(\ref{eq:10}) and
Eq.(\ref{eq:11}) into the $S$ matrix, we have the following standard form,
\begin{eqnarray}
S=\left(\begin{array}{cc}
  rI &adj(T) \\
 T& r'I \\
\end{array}\right)\xrightarrow{\text{if}  \det(T)\neq 0} \left(\begin{array}{cc}
  rI & tU^{-1} \\
  tU& r'I \\
\end{array}\right) \label{standardgeneral} 
\end{eqnarray}

Now let's prove if $\det(T)\neq 0$, then $\det(U)=1$. Suppose $\det(U)=-1$, then, the
eigenvalues are $\lambda$ and $-\frac{1}{\lambda}$. The
corresponding eigenvalues of $U^{-1}$ must be $\frac{1}{\lambda}$
and $-\lambda$. Since the eigenvalues of $U$ should coincide with
that of $U^{-1}$, we have $\lambda=\pm 1$. So, we have the general
form  $U=\left(\begin{array}{cc}
 s & a\\
  b& -s \\
\end{array}\right)$ with $\sqrt{s^2+ab}=1$. From Eq.(\ref{eq:11}), we have similar form
for $U^{-1}$, which reads, $\left(\begin{array}{cc}
 -s & -a\\
  -b& s \\
\end{array}\right)$. Now, from $UU^{-1}=1$, we have $-s^2-ab=1$,
which is in contradiction with the presumption that
$\det(U)=-s^2-ab=-1$. So, we have $\det(U)=1$. \emph{Q.E.D}

Here it's noteworthy to point out that although we assumed on the above the simplest scattering setup with two mono-channel leads,  the procedure can  be easily applied to the
general  scattering problem with multiple leads attached to the central region, each with multiple propagating channels.
 Only to note that for the general case, the standard form Eq.(\ref{standardgeneral}) should be understood
as the scattering sub-matrix between arbitrary two spinful channels.

Now let's further prove that in the two-terminal mono-channel case, $U$ is an
unitary matrix. 
From the unitarity property of $S$, we get
$tr^*U+r't^*U^{-1\dagger}=0$. By multiplying $U^{-1} $ from the
right side, we have $tr^*+r't^*U^{-1\dagger}U^{-1}=0$, from which,
\begin{eqnarray}
U^{-1\dagger}U^{-1}=eI.
\end{eqnarray}

Since $\det U^{-1}=1$ and $TrU^{-1\dagger}U^{-1}> 0$ , we get $e=1$. So, $U$ is nothing but a $SU(2)$ matrix. In addition, we have $r'=-\frac{t}{t^*}r^*$. 

 The property that the 
scattering sub-matrices has the standard form Eq.(\ref{standardgeneral}) with
$\det U=1$ for the general case and $U\in SU(2)$ for the simplest case with two mono-channels constitutes the central result of this paper. In the following we will make some comments.
 


Firstly, we would prove that the analytical property of the scattering matrix agree well with the polar decomposition widely used in the random matrix community.  

In the two-terminal case with mono-channel leads, according to the polar decomposition\cite{Beenakker}, we can get the following neat form:
\begin{eqnarray}
S=\left(\begin{array}{cc}
  -\sqrt{1-P}e^{i\theta_1} & \sqrt{P}e^{i\phi}\Omega \\
  \sqrt{P}e^{i\phi}\Omega^{\dagger}& \sqrt{1-P}e^{i\theta_2} \\
\end{array}\right)   \label{polardecomp} 
\end{eqnarray}
where $P$ is the transmission probability, $\phi=\frac{\theta_1+\theta_2}{2}$, and $\Omega$ is a SU(2) matrix. Clearly, this form is a specific example of the standard form Eq.(\ref{standardgeneral}) with $\det(T)\neq 0$   discussed above.
 Actually, under basis transformation between normal basis, the scattering matrices with the same $P$ but with different $\theta's$ and $\Omega$ of a two-channel  problem can be transformed into each other. Let's investigate this point in more detail as follows.

For the first step,  let's identify the general basis transformation for the propogating modes which keeps the time reversal operator  $\Theta$ invariant. Consider a basis transformation $g$ for right propogating modes and $h$ for left propogating modes of the left lead. By definition, we have,
\begin{eqnarray}
|1\sigma\rangle_{new}&=&\Sigma_{\sigma'}g_{\sigma,\sigma'}|1\sigma'\rangle\nonumber\\
|-1\sigma\rangle_{new}&=&\Sigma_{\sigma'}f_{\sigma,\sigma'}|-1\sigma'\rangle
\end{eqnarray}

 If the transformation matrices $f$ and $g$ satisfy,
\begin{eqnarray}
f=\left(\begin{array}{cc}
  g_{22}^*&-g_{21}^* \\
  -g_{12}^*& g_{11}^* \\
\end{array}\right)   \label{fg} 
\end{eqnarray}
it's straitforward to check that $\Theta|\pm 1\sigma\rangle_{new}=\sigma|\mp 1\bar{\sigma}\rangle_{new}$ (in accordance with  Eq.(\ref{eq7})) which  means $\Theta$ is invariant under the basis transformation. 

Now, if we have transformation matrices $G_L^*(F_L^*)$ and  $G_R^*(F_R^*)$ for incoming(outgoing) basis states of left and right leads, respectively, defining block matrices  
$
G=\left(\begin{array}{cc}
  G_L&0 \\
  0& G_R
\end{array}\right)  
$
 and 
$
F=\left(\begin{array}{cc}
  F_L&0 \\
  0& F_R
\end{array}\right)  
$, the wave amplitude vectors transform as $a_{new}=Ga$ and  $b_{new}=Fb$.  Then, from $b_{new}=S_{new}a_{new}$,  we get the transformation of scattering matrix,
\begin{eqnarray}
 S_{new}=FSG^{\dagger} \label{Smatrixtransform}
\end{eqnarray}
    
Let's consider two kinds of typical transformations starting from the standard form Eq.(\ref{standardgeneral}) with $\det(T)\neq 0$. Firstly, let's take $F_L=G_L^*=e^{i\theta_L}I$ and $F_R=G_R^*=e^{i\theta_R}I$, it's straitforward to see,
\begin{eqnarray}
r_{new}&=&re^{2i\theta_L},r_{new}'=r'e^{2i\theta_R},\\
t_{new}&=&te^{i(\theta_L+\theta_R)}  \label{rtnew} 
\end{eqnarray}
from which one sees that the phase of $t$ can be changed. So, we can choose $t_{new}=\sqrt P e^{i\phi}$. Furthermore, let's assume $r_{new}=-\sqrt{1-P}e^{i\theta}$, so that $r'_{new}=\sqrt{1-P}e^{i(2\phi-\theta)}$ due to unitarity. 
Then, let's choose $\theta_L=-\theta_R=\frac{\theta_1-\theta}{2}$ and do transformation Eq.(\ref{rtnew}) once more , we can get the form Eq.(\ref{polardecomp}) with $U$ instead of $\Omega$.


Secondly, we choose  $F_{L(R)}=G_{L(R)}=O_{L(R)}$ where $O_{L(R)}\in SU(2)$, then the scattering matrix keeps its standard form and transforms as,
\begin{eqnarray}
U_{new}=O_LUO_{R}^{\dagger} \label{rtnew3} 
\end{eqnarray}
with other coefficients $t$, $r$,$r'$ invariant. For the case $O_L=O_R=O$, the $SU(2)$ basis transformation is nothing but the spin rotation operator for the whole system. We can always find an $O$ to diagonalize $U$, which is unitary. If the whole system has spin rotational symmetry, $U=\pm I$(note $t$ is also undetermined up to a minus sign), thus in any spin-rotated frame the scattering matrix is the same.

 Combining the two kinds of transformations that keep transmission probability $P$ invariant, with proper $\theta_L$, $\theta_R$ and $O$, we can transform any scattering matrix with standard  form into a form of Eq.(\ref{polardecomp})(One can even fix $\phi=0$ and $\Omega=1$ for target).  Thus we've completed the discussion of basis transformations for the case $\det T\neq 0$. In essence, we've demonstrated the mutual equivalence of  two terminal mono-channel scattering problems which has the same $P$. 


  It's tempting to assume that the geneneral standard form is redundant since the essential thing in the scattering matrix is $\sqrt{P}$ in Eq.(\ref{polardecomp}). This is correct if all we concern about is  static transport properties. However, if we deal with the time-dependent transport problem, such as the quantum pumping problem, the phase information in scattering matrix will also become essential.  One example is the proposal of pure spin current
 generation discussed in Ref\cite{Nagaosa}, where $U$ is assumed to be the most simple form, i.e.,
 a  diagonal $SU(2)$ matrix. It would be a straitforward extension if we consider a general $SU(2)$ matrix for $U$. 


The polar decomposition is widely used in the random matrix theory for mesoscopic transport\cite{Beenakker}. The unitary transformations for incoming states and outgoing states satysfying Eq.(\ref{fg}) is mutually dual in the language of quaternion matrix\cite{Beenakker}. On the above we proved the equivalence of the standard form  and the polar decomposition form for the two-terminal mono-channel case.  
Taking into account that the transformation matrix $G$ and $F$ are dual quaternion matrices, we can even prove  for  multi-channel case (through some lengthy but routine steps) that the polar decomposition always respects the standard form.  The anlytical structure of the standard form, i.e., Eq.(\ref{standardgeneral}) together with (1)$\det T=0$, or, (2)$\det T\neq 0$ and $\det U=1$, however, can be applied to the most general multi-terminal multi-channel case. We believe the manifest symmetrical structure of the sub-matrices of scattering matrix is less noticed before and is worthy of a new report here.  
Finally, we comment that following the direct procedure we adopted in this short report, the standard form of the scattering matrices subject to various other symmetry constraints(see,e.g., Ref\cite{Akhmerow}) can be similarly deduced.
\begin{acknowledgments}
We are grateful to C. W. J. Beenakker for drawing our attention to the relationship of standard form given in Eq.(\ref{standardgeneral}) and the the polar decomposition of  the scattering matrix.  We acknowledge the financial support from the National
Natural Science Foundation of China (under Grants
No. 11004174 (Y. J. Jiang) and 11174252 (F. Zhai)).
\end{acknowledgments}

\end{document}